\documentstyle[prl,twocolumn,aps,epsf]{revtex}
\begin{document}
\draft
\twocolumn
\title{
Intrinsic frustration effects in anisotropic superconductors}
\author{F. Guinea}
\address{
Instituto de Ciencia de Materiales.
Consejo Superior de Investigaciones Cient{\'\i}ficas.
Cantoblanco. 28049 Madrid. Spain.  \\}
\date{\today}
\maketitle 
 
\begin{abstract}
Lattice distortions in which the axes
are locally rotated provide an intrinsic source of frustration
in anisotropic superconductors. A general framework to study
this effect is presented. The influence of lattice defects
and phonons in $d$ and $s+d$ layered superconductors is
studied.
\end{abstract}

\pacs{73.40.Gk,74.25.Bt,74.72.-h}
\narrowtext
\section{Introduction.}
In anisotropic superconductors, local rotations of the lattice
modulate the order parameter. The most striking manifestation
of this effect is the influence of grain boundaries in high-T$_{\rm c}$
superconductors\cite{T94,K95,SR95}. 
Less pronounced effects are also to be expected from
lattice defects such as dislocations, as they twist
the lattice axes in their vicinity. Transversely polarized
phonons can, in principle, also couple to the superconducting order
parameter through the local rotations that they induce.

If, along a given direction, there are strains which change
the orientation of the lattice axes, the equilibrium order parameter must
follow that distortion. For instance, in a two
dimensional square lattice, a parameter with $d_{x^2-y^2}$ 
symmetry, $\Delta_{x^2-y^2}$, must change into 
$\cos ( 2 \theta ) \Delta_{x^2-y^2} - \sin ( 2 \theta ) \Delta_{xy}$
after a lattice rotation by an angle $\theta$. We express the local
order parameter in terms of its components in a fixed frame of
reference which is independent of the orientation of the lattice.
We expand the local order parameter in terms of the functions
$( \Delta_{x^2-y^2} , \Delta_{xy} )$, defined in this external frame.
In principle, the symmetry of the square lattice allows for mixing of
a pure $\Delta_{x^2-y^2} = \Delta_0 \cos ( 2 \theta )$ function with
higher spherical harmonics. Under a rotation, each of these 
harmonics behaves differently, and requires a specific description
in the external frame of reference mentioned earlier. For simplicity,
we will first ignore these higher order components. Their inclusion
will be discussed later, as well as extensions to more complex 
order parameters.

\section{Formalism.}

We will now set up an effective Ginzburg Landau description of 
the situation described above. Let us first consider a
model superconductor with very small coherence length.
At each position, the order parameter will match perfectly the 
orientation of the lattice axes. There is no cost in free energy
due to misorientations in the order parameter. We can ignore
gradient terms, and use an expression which contains only
quartic and quadratic terms, provided that we insert in them
the value of order parameter with respect to the local
axes.

The coupling between neighboring regions is described by gradient terms.
There is a free energy cost associated to the inhomegeneities of
the order parameter. In the present case, the degree of inhomegeneity
must be referred to the fixed, external frame of reference. 
There is an increase in the free energy of the system when different
areas of the lattice are misoriented, even if the local order
parameter is aligned with the respective axes. A particular example of this 
situation is a grain boundary, as described in the introduction.

The difference between the two ways of describing the order parameter,
with respect to an external frame or to the local one, is determined
by the degree of rotation of the lattice from one region to another.
Let us define the rotation which takes the axes from one point
to another by the angle $\theta$. As mentioned previously, an
order parameter which follows the lattice will look,
in the external frame, like
$\Delta_{x^2-y^2}$ in one point and $\cos (2 \theta ) \Delta_{x^2-y^2}
+ \sin ( 2 \theta ) \Delta_{xy}$ in the other.
The transformation which takes one expression into the other can be
accomplished by modifying  the two component vector 
$( \Delta_{x^2-y^2} , \Delta_{xy} )$ by means of an operator of the
type $e^{i \int {\bf \vec{A}} d
\vec{r}}$, such that: 
\begin{equation}
i \int {\bf \vec{A}}_g d \vec{r} = 2 \theta \left( 
\begin{array}{cc} 0 &1 \\ -1 &0 \end{array} \right)
\end{equation}
and:
\begin{equation}
e^{i \int {\bf \vec{A}}_g d \vec{r}} = \left( \begin{array}{cc}
\cos ( 2 \theta ) &\sin ( 2 \theta ) \\ - \sin ( 2 \theta )
&\cos ( 2 \theta ) \end{array} \right)
\end{equation}
The integral is to be taken along the path in real space
along which the lattice is rotated. 
As $\theta$ is the total rotation along
the path considered, we can write, for the example considered here:
\begin{equation}
{\bf \vec{A}}_g = 2 ( \nabla {\bf R} ) \sigma_y = {\bf \vec{A}} \sigma_y
\label{gauge}
\end{equation}
where $\nabla$ is the gradient operator in real space, ${\bf R}$
specifies the local rotation of the lattice axes,
and $\sigma_y$ is a Pauli matrix. The rotation
${\bf R}$ can be written as\cite{LL59}:
\begin{equation}
{\bf R} = \frac{\partial_x u_y - \partial_y u_x}{2}
\label{rotation}
\end{equation}
where ${\bf \vec{u} ( \vec{r} )}$ are the local deviations 
of the lattice node at ${\bf \vec{r}}$ from equilibrium.
Eq. (\ref{rotation}) is valid for small deviations (strains).

The expressions above allow us to relate the changes in the order
parameter as seen in the two frames, and the rotations of the lattice,
in terms of the local strains. The gradient term in the Ginzburg
Landau description, which should be expressed with respect to the
external frame, looks, in terms of the order parameter with respect
to the local axes:
\begin{eqnarray}
{\cal F} &= 
&\frac{a_{x^2-y^2}
 \bar{T}_c \xi_0^2}{2} \int \sum_{i=x,y} | ( \partial_i - A_{gi} ) 
\Delta_{x^2-y^2} |^2 \nonumber \\ &+
&\frac{a_{xy}
 \bar{T}_c \xi_0^2}{2} \int \sum_{i=x,y} | ( \partial_i - A_{gi} ) 
\Delta_{xy} |^2 \nonumber \\ 
&+ &\frac{a_0 ( T - T^{x^2-y^2}_c )}{2} \int | \Delta_{x^2-y^2} |^2 
\nonumber \\ &+
&\frac{a_0 ( T - T^{xy}_c )}{2} \int 
| \Delta_{xy} |^2 + {\rm quartic \, \, terms}
\label{GL}
\end{eqnarray}
where ${\cal F}$ is the free energy.
$\bar{T}_c$ is a number
with dimensions of temperature, and of order $T^{x^2-y^2}_c$.
Finally, $a_{x^2-y^2} \sim a_{xy} \sim a_0$ are proportionality constants
which play no role in the analysis presented below.
We take the critical temperatures of the $\Delta_{x^2-y^2} ,
\Delta_{xy}$ order parameters as different, as appropiate for
a square lattice. Equation (\ref{GL}) and the definitions (\ref{gauge})
and (\ref{rotation}) suffice to study the phenomenology of layered
superconductors whose order parameter is well approximated by 
$[ \cos ( 2 \theta ) , \sin ( 2 \theta ) ]$. 

It is easy to see that, when $\Delta_{x^2-y^2}$ and $\Delta_{xy}$
have the same critical temperature, and $\nabla \times {\bf \vec{A}} = 0$,
the effects of the field can be made to vanish by performing
a local rotation of the order parameter\cite{disc}. 
In this case, the
system has isotropic superconducting properties. Hence, rotations
of the underlying crystal lattice leave the (degenerate) order parameter
unaffected.

The formalism described here can be expressed in terms of a
connexion, which depends on the lattice orientation.
The definition of parallel transport of a vector,
in our case $(\Delta_{x^2-y^2} , \Delta_{xy} )$, needs to be
modified by the rotation of the lattice. The usual derivative
is changed into a covariant derivative\cite{N90}. 
This technique lies at the basis
of the extensive work done in topological phases\cite{SW88},
and has found use in many fields in condensed matter physics\cite{T95}.

In addition to
the effects described by (\ref{GL}), there may be
an explicit coupling of the order parameter
to the lattice strains. The simplest
coupling is  proportional to $ \frac{\partial T_c}
{\partial P} \int \sum_i \epsilon_{ii} |
\Delta_{x^2-y^2} |^2$, where $\sum_i \epsilon_{ii}$
defines the local compression, or expansion, in terms
of the symmetric part of the strain tensor. This term
can also exist in a $s$ wave superonductor. A coupling of this type
is not negligible in the high-T$_{\rm c}$ compounds,
due to the strong dependence of T$_{\rm c}$ on pressure\cite{G88}.

The previous analysis can be generalized to three dimensional
superconductors with arbitrary order parameters. Local rotations
of the lattice can be expressed in terms of the parameter:
\begin{equation}
R_i = \frac{\epsilon_{ijk} \partial_j u_k}{2} 
\label{rotation3d}
\end{equation}
where $\epsilon_{ijk}$ is the fully antisymmetric tensor.
In order to determine the transformations 
induced by these rotations in the order parameter, we must first
decompose it into spherical harmonics with well defined transformation
properties. Then, a gauge field which performs rotations in
order parameter space is added to each component. Finally, as in
(\ref{GL}), all terms allowed by the lattice symmetry are included
in the free energy.
The procedure is cumbersome, but straightforward.
For instance, we can add to (\ref{GL}) higher harmonics
which transform like $\cos [ ( 4 n + 2 ) \theta ] , \sin
[ ( 4 n + 2 ) \theta ]$. The gauge field associated to
these components is of the type ${\bf \vec{A}}_g = 4 ( 2 n + 1 ) 
\sigma^n_y \nabla ( \partial_x u_y - \partial_y u_x )$,
where $\sigma^n_y$ transforms one of these parameters
into the other. 
The gauge field induced by the higher
harmonics of the order parameter are also proportional
to the gradient of the lattice rotations, but with a larger
constant of proportionality. The term with $|{\bf {\vec A}}_g|^2$
in the free energy induces a reduction of these components of
the order parameter. This reduction will be stronger for higher harmonics than
for lower ones.

\section{Static distortions.}

We now consider specific lattice distortions within the layers.
The simplest static defect is a dislocation. The strains
in the vicinity of a dislocation can be calculated from
the theory of elasticity. If the Burgers vector is
${\bf \vec{b}}$, then\cite{LL59}, 
\begin{equation}
\frac{\partial_x u_y - \partial_y u_x}{2}
= \frac{{\bf \vec{b} \cdot \vec{r}}}{ \pi |{\bf \vec{r}}|^2}
\end{equation}
where we are neglecting  a possible anisotropy in the elastic constants
of the lattice.
We now assume that the $\Delta_{xy}$ component is negligible,
which is justified if, for instance, the parameter $T_c^{xy}$ in
(\ref{gauge}) is less than zero. Then, the coupling of the
lattice rotations to $\Delta_{x^2-y^2}$ is due to terms quadratic
in the gauge potential, and we obtain:
\begin{equation}
\frac{a_{xy} \bar{T}_c \xi_0^2}{2} \int | \Delta_{x^2-y^2} |^2
| {\bf \vec{A}} |^2
\label{coupling2}
\end{equation}
and $| {\bf \vec{A}} |^2 = \frac{ 
| {\bf \vec{b}} |^2}{2 \pi^2 | {\bf \vec{r}} |^4}$.
The effect of this term is to reduce the effective critical temperature
near the dislocation. It suppresses a possible local 
enhancement of T$_{\rm c}$ due to the explicit dependence
of the critical temperature on pressure\cite{G88}. In the
presence of a magnetic field, the normal cores of vortices
will be attracted towards the dislocations, to take
adventage of the supression of T$_{\rm c}$, leading to pinning.
Near the core of the dislocation, $r \sim \xi_0$, we can estimate the
reduction of T$_{\rm c}$ (see below) as $\frac{\Delta T_c}{T_c}
\sim \frac{b^2}{\xi_0^2}$. This effect can be significant in a
material with a short coherence length, such as the copper
oxides.

A grain boundary can be modelled as an array of dislocations.
The gauge fields induced by each of them can be added, to obtain the 
effect of the whole boundary. After some algebra, we find that,
if the rotation of the axes across the boundary is not large:
\begin{eqnarray}
| {\bf \vec{A}} |^2 &= &\frac{\pi^2 \sin^4 ( \theta )}{2 a^2}
\frac{ \left[ 1 - \cos \left( \frac{\pi y}{\bar{a}} \right) \cosh
\left( \frac{\pi x}{\bar{a}} \right) \right]^2 +
\sin^2 \left( \frac{\pi y}{\bar{a}} \right)}{\left[ \cosh \left(
\frac{\pi x}{\bar{a}} \right) - \cos \left( \frac{\pi y}{\bar{a}} \right)
\right]^4} \nonumber \\ &\sim
&\frac{\pi^2 \sin^4 ( \theta )}{4 a^2} e^{- \frac{ \pi x}{\bar{a}}} ,
\, \, \, \, \, x \rightarrow \infty
\label{grainb}
\end{eqnarray}
where $a$ is the lattice constant, $\theta$ is the angle between
the axes at both sides of the boundary,
and $\bar{a} = \frac{a}{\sin ( \theta )}$. From (\ref{grainb}) and (\ref{GL}),
we can estimate the reduction of the critical temperature at the boundary.
We assume that $a_{x^2-y^2} \sim a_{xy} \sim a_0$, 
and $\bar{T}_c \sim T_c^{x^2-y^2}$. Then,
$\frac{\Delta T}{T_c} \sim \frac{\pi^2 \xi_0^2 \sin^4 ( \theta )}{2 a^2}$,
where $T_c = T_c^{x^2-y^2}$.
This calculation is rigorously valid when $\sin ( \theta ) \ll 1$,
that is, when strains change slowly. 
A significant reduction of T$_{\rm c}$ can be expected near
a twin boundary, which will act in a way similar to a weak link.
 It would be interesting 
to check this experimentally\cite{K97}.

\section{Dynamic rotations.}

The orientation of the lattice can also be modified dynamically, 
by transversely polarized phonons.  A longitudinal phonon
induces local compressions and dilatations of the lattice.
A transverse phonon does not change the volume of the
unit cell, but induces a shear deformation which changes
from position to position. This deformation can be viewed
as a local rotation of the axes. In a material with a
complex unit cell, there are also optical phonons
which rotate locally the axes. The rotations can be analyzed
by means of the formalism discussed before. The only difference
with a static distortion is that the gauge field which
accounts for the lattice rotations depends on dynamic
displacements. The time scales associated to the
superconducting response are of the order of the inverse plasma
frequency. This frequency is much higher than typical
phonon frequencies, so that the changes in T$_{\bf c}$
induced by phonons can be studied as if the lattice distortions
were static.

Transversely polarized acoustic phonons give rise to strains proportional
to the gradient of their amplitudes. Hence,
we expect that the Fourier components of ${\bf R}$ are given by:
\begin{equation} 
{\bf R}_k \sim {\bf \vec{k}}_{\perp} 
\times {\bf \vec{u}}_k
\end{equation}
where ${\bf \vec{u}}_k$ is the amplitude of
a phonon of momentum ${\bf \vec{k}}$, and $k_{\perp}$ is
the component of ${\bf \vec{k}}$ perpendicular to ${\bf \vec{u}}_k$.
For transversely polarized phonons, ${\bf \vec{k}}$
is always perpendicular to ${\bf \vec{u}}_k$, and 
we do not need to take this restriction into account.

The gauge potential
in (\ref{gauge}) has a finite average, related to the thermal
(or quantum) average value of the phonon amplitudes. It is given by:
\begin{equation}
|{\bf \vec{A}}_g|^2 \sim \sum_k \langle  {\bf \vec{k}}^2  
| ( {\bf \vec{k}} \times {\bf \vec{u}}_k ) |^2 \rangle
\label{gaugeph}
\end{equation}

We are analyzing materials which show strong differences
in the superconducting properties along the in plane and out of plane 
directions. Phonons, however, need not be so anisotropic. We
assume that they are described by an isotropic sound velocity,
$c$, and a Debye frequency $\omega_D \approx \frac{\hbar c}{a}$,
where $a$ is the lattice constant, which we take to be of
the same order of magnitude in all directions. 
The three dimensional integral implicit in (\ref{gaugeph}) is
extended to those modes such that $\hbar c k \leq k_B T$
and $a \leq k_z^{-1}$ and $\xi_0 \leq k_{\parallel}^{-1}$.
Note that the Ginzburg-Landau description used here
is valid at length scales greater than $\xi_0$\cite{phononsw}.
For simplicity, we take the Bose occupancy of a mode 
of frequency $\omega$ to be $\frac{k_B T}{\hbar \omega}$.
None of these approximations will alter the order of 
magnitude of the effect. 
We can distinguish three regimes: i) if $k_B T \ll \hbar c \xi_0^{-1}$,
the upper cutoff in (\ref{gaugeph}) is due to the Bose factor.
ii) when $\hbar c \xi_0^{-1} \ll k_B T \ll \hbar c a^{-1}$,
the upper cutoff is $k_{\parallel} \sim \xi_0^{-1}$ and 
$k_z \sim \frac{k_B T}{\hbar c}$ and iii) if $\hbar c a^{-1}
\ll k_B T$, the maximum value of $k_z$ is $a^{-1}$.
We can now estimate the reduction in the (average) value of the
critical temperature in the same way as in the previous section. We find:
\begin{equation}
\frac{\Delta {\rm T_c}}{{\rm T_c}} \sim \left\{ \begin{array}{lr}
\frac{\xi_0^2}{a^2} \frac{\hbar}{\rho
a^3 \omega_D} \left( \frac{k_B {\rm T_c} }{\hbar \omega_D } \right)^6 
&k_B {\rm T_c} \ll \frac{\hbar \omega_D a}{\xi_0} \\  
\frac{\hbar}{\rho a^3 \omega_D} 
\left( \frac{k_B {\rm T_c}}{\hbar \omega_D} \right)^4
&\frac{\hbar \omega_D a}{\xi_0} \ll k_B {\rm T_c} \ll
\hbar \omega_D \\ 
\frac{\hbar}{\rho a^3 \omega_D} 
\frac{k_B {\rm T_c}}{\hbar \omega_D}  &\hbar \omega_D \ll k_B {\rm T_c}
\end{array}
\right.
\label{Tcphonons}
\end{equation}
where $\rho$ is the density.
This reduction of the
critical temperature should not be negligible in the 
high-T$_{\rm c}$ compounds, as $\xi_0$ is not very different
from $a$, the lattice constant.

In addition to a reduction in the average value of T$_{\rm c}$,
we can also analyze the fluctuations in this value. We are treating
phonons as a source of static disorder, which reduces T$_{\rm c}$.
The value of ${\bf \vec{A}}_g$ fluctuates, changing the local
values of the critical temperature. If these fluctuations take place
on scales larger than $\xi_0$, a distribution of T$_{\rm c}$'s will
be observed. As the changes in T$_{\rm c}$ go as:
\begin{equation}
\Delta {\rm T_c} \sim {\bf \vec{A}}_g^2 \sim
\sum | k |^4 n_k |{\bf \vec{u}}_k |^2
\end{equation}
we find that:
\begin{equation}
\langle \Delta {\rm T_c} ( {\bf \vec{r}} ) \Delta {\rm T_c}
( {\bf \vec{r}}' )  \rangle  \sim
\sum {\bf \vec{k}}^8 
e^{i {\bf \vec{k}} ( {\bf \vec{r}}
- {\bf \vec{r}}' )} | {\bf \vec{u}}_k |^4
\label{fluctuations}
\end{equation} 
and:
\begin{equation}
\langle \Delta {\rm T_c}^2 \rangle 
\sim \left\{ \begin{array}{lr}
{\rm T_c}^9 &k_B {\rm T_c} \ll \frac{\hbar \omega_D a}{\xi_0} \\ 
{\rm T_c}^7 &\frac{\hbar \omega_D a}{\xi_0} \ll k_B {\rm T_c} \ll
\hbar \omega_D \\ 
{\rm T_c} &\hbar \omega_D \ll k_B {\rm T_c} \end{array}
\right.
\label{fluctuationsTc}
\end{equation}
by comparing (\ref{Tcphonons}) and this expression, one can see
that fluctuations cannot be neglected. Finally, the typical 
length scale for fluctuations inside the planes is 
\begin{equation}
\xi \sim {\rm max} \left( \frac{\hbar c}{k_B {\rm T_c}} , \xi_0 \right)
\sim {\rm max} \left( \frac{\hbar \theta_D}{k_B {\rm T_c}} ,
\frac{W}{k_B {\rm T_c}} \right) \sim \xi_0
\end{equation}
where $W$ is the electronic bandwidth. Hence, we expect large fluctuations 
in small scale domains. Transport measurements will be unable to
resolve this structure, and the average ${\rm T_c}$ will be observed.

Transverse optical phonons of frequency $\omega_{opt}$.
give an additional reduction of T$_{\rm c}$,
\begin{eqnarray}
\frac{\Delta T_c}{T_c} = \left\{ \begin{array}{lr} 
\frac{\langle u^2 \rangle}{a^2} 
e^{-\frac{\hbar \omega_{opt}}{k_B T_c}}
&\hbar \omega_{opt} \ll k_B T_c \\ \frac{\langle u^2 \rangle}
{a^2}\frac{k_B T_c}{\hbar \omega_{opt}} 
&k_B T_c \gg \hbar 
\omega_{opt} \end{array} \right.
\label{phononopt}
\end{eqnarray}
where $\langle u^2 \rangle$ denotes the mean square displacement 
of the phonon. It is, typically, a fraction, $10^{-3} - 10^{-1}$
of $a^2$. As in the previous case, we assume that the superconducting
order parameter has a short coherence length in the out of plane direction.
There is a contribution like (\ref{phononopt}) for each 
optical phonon transversely polarized in the plane.

The frustration of the order parameter in anisotropic superconductors
by dynamical vibrations implies that $T_c$ cannot be much larger 
than the frequency of the phonons responsible of the effect.

\section{Coupling between distortions and currents.}

The Ginzburg Landau expression (\ref{GL}) contains terms which are linear
in ${\bf \vec{A}}_g$, which couple one component of the order parameter
and the spatial derivatives of the other
(note that ${\bf \vec{A}}_g$ is an operator
which rotates one component into the other). These derivatives are different
from zero in the presence of a supercurrent, because of the variation
in the phase. In the case described by (\ref{GL}), however, these terms
do not contribute, as $\Delta_{xy} \sim 0$.

Terms of this type play a role in
materials where the superconducting properties differ
significantly along the lattice axes.
Let us consider an orthorhombic superconductor,
where the order parameter is $s+d$. 
In addition to the terms in (\ref{GL}) we also have:
\begin{eqnarray}
\Delta {\cal F}_{ortho} &= &\frac{\bar{a}' \bar{T}_c \xi_0^2}{2}
\int \partial_x \Delta_s \partial_x \Delta_{x^2-y^2}^* -
\partial_y \Delta_s \partial_y \Delta_{x^2-y^2}^*  \nonumber \\
&+ &\frac{\bar{a}' \bar{T}_c \xi_0^2}{2}
\int \partial_x \Delta_s \partial_y \Delta_{xy}^* + \partial_y
\Delta_s \partial_x \Delta_{xy}^* + c. c. \nonumber \\ 
&+ &\frac{a_s \bar{T}_c \bar{\xi}_0^2}{2} \int \sum_{i=x,y}
| \partial_i \Delta_s |^2 + \frac{a_s ( T - T_c^s )}{2} \int
| \Delta_s |^2 \nonumber \\ 
&+ &\frac{a_{sd} ( T - {T'}^* )}{2} \int \Delta_s \Delta_{x^2-y^2}^*
\nonumber \\ &+ &c. c. + {\rm quartic \, \, terms}
\label{ortho}
\end{eqnarray}
where $\Delta_s$ is the $s$ component of the order parameter.
For the sake of simplicity, we ignore terms induced by the
orthorhombic symmetry which are quadratic in the gradients\cite{gradients}.
These terms, needed for a complete description of the anisotropy in
the penetration depth, do
not modify qualitatively the picture
discussed in the previous sections.

In the absence of internal lattice rotations, $\Delta_{xy} = 0$. 
The first term in (\ref{ortho}) leads to a difference in
the penetration depth along the two lattice axes.
Local rotations of the lattice can be incorporated
by adding the gauge field (\ref{gauge}) to the derivatives 
of the $d$ components of the order parameter in (\ref{ortho}).
Then, assuming that the equilibrium order parameter  has both
$\Delta_{x^2-y^2}$ and $\Delta_s$ components, we obtain terms of the type:
\begin{equation}
\frac{\bar{a} \bar{T}_c \xi_0^2}{2} \int \partial_x
\Delta_s A_y \Delta_{x^2-y^2}^* + \partial_y \Delta_s A_x
\Delta_{x^2-y^2}^* + c. c.
\label{orthoder}
\end{equation}
This expression gives a contribution if $\nabla \Delta_s \ne 0$,
and, in particular, in the presence of a current. 

The changes in a current carrying state induced by lattice rotations
are best studied within London's framework, which
is valid if the magnitude of the order parameter does not change.
We define the London tensor, with
the same symmetry properties of the perfect lattice, such that
${\bf \vec{j}} = {\bf m^{-1} \vec{A}}_{elec}$,
where ${\bf \vec{A}}_{elec}$ is the electromagnetic vector
potential.
For the study
of currents within the layers, we have:
\begin{equation}
{\bf m}^{-1} \equiv c \left( \begin{array}{cc} \lambda_x^{-2} &0 \\ 
0 &\lambda_y^{-2} \end{array} \right) 
\label{londont}
\end{equation}
where $\lambda_x , \lambda_y$ are the penetration depths along the
two axes, and $c$ is the velocity of light. 
London's expression for the free energy is\cite{dG66}:
\begin{equation}
{\cal F}_L = \frac{1}{8 \pi} \int | \nabla \times{\bf \vec{A}}_{elec} |^2 +
\frac{\pi}{c^2} \int \lambda_x^2 \frac{j_x^2}{2} + 
\lambda_y^2 \frac{j_y^2}{2}
\label{london}
\end{equation}
The deviation of tetragonal symmetry is given by $\lambda_x^{-2}
- \lambda_y^{-2} \propto \frac{\bar{a}' T_c \xi_0^2}{2} | \Delta_s
\Delta_{x^2-y^2}^* |$.

\begin{figure}
\begin{center}
\mbox{\epsfxsize 8cm\epsfbox{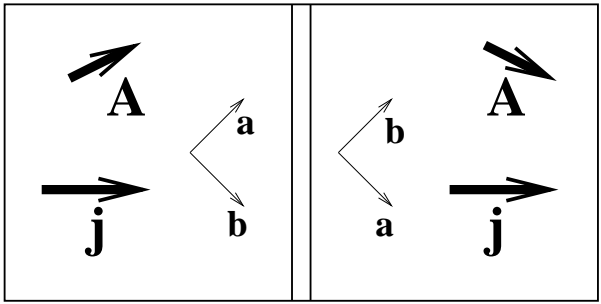}}
\end{center}
\caption{Currents and electromagnetic vector potentials near a twin boundary,
in a slab much narrower than the penetration depth. When the current is
constant, and flows parallel to the boundaries, the vector potential must
have a discontinuity at the twin. This discontinuity induces a magnetic
field in the direction perpendicular to the slab.}
\label{twin}
\end{figure}

Taking into account that the gauge field related to the lattice distortions,
${\bf \vec{A}}$, is itself a derivative, we
can integrate by parts in eq.(\ref{orthoder}), to obtain:
\begin{eqnarray}
\Delta {\cal F}_L &= &\frac{\bar{a} T_c \xi_0^2}{2}
\int \partial_x \Delta_s {\bf R} \partial_y \Delta_{x^2-y^2}^*
\nonumber \\ &\propto &( \lambda_x^{-2} - \lambda_y^{-2} ) 
\lambda_x^2 \lambda_y^2 \int j_x ( \partial_x u_y
- \partial_y u_x ) j_y
\label{londonr}
\end{eqnarray}
This equation can be interpreted as a rotation of the London tensor,
(\ref{londont}), induced by the distortion of the lattice.
The currents adjust to the directions of the local principal axes.
At a grain boundary, for instance, ${\bf \vec{A}}_{elec}$ 
must be continuous. The component of the current parallel to
the boundary must be discontinous, leading to a kink in the
direction of the total current, as the London tensor which
gives the current at both sides is 
discontinuous\cite{G89,G90,FBBK97}. 
If, on the other hand, the boundary
conditions prevent the current from acquiring a transverse component,
the vector potential shows a kink, and a perpendicular magnetic field
is generated at the boundary, as sketched in fig(\ref{twin}\cite{twinb}.
This picture, derived from London's theory, is valid at distances
comparable to the penetration depth. The discontinuity at the
boundary is rounded off at shorter scales.

\begin{figure}
\begin{center}
\mbox{\epsfxsize 8cm\epsfbox{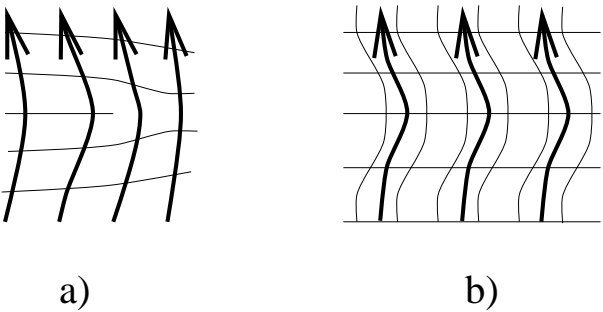}}
\end{center}
\caption{Sketch of the a currents flowing near a dislocation (a),
and instantaneous current in the presence of a transverse acoustic phonon (b).}
\label{currents}
\end{figure}

The present analysis implies that currents can also be deflected by other
types of static defects, and by phonons. The currents flowing near
a dislocation, and 
the instantaneous current in the presence of phonons, are sketched in 
fig.(\ref{currents}). Magnetic fields will be induced. It is interesting
to note that phonons can modulate the currents and give rise to
electromagnetic fields. 

So far, we have only considered the role of in plane distortions
in layered superconductors. If we include rotations around in plane
axes, we must allow for components like $\Delta_{3z^2-r^2} , \Delta_{xz} ,
\Delta_{yz}$ in the order parameter.
Because of the strong anisotropy of these compounds, components which
transform like higher powers of $z^2$ are also to be expected.
In analogy with (\ref{london}), these components lead to contributions
of the type $\int j_i ( \partial_i u_z - \partial_z u_i ) j_z$,
where the direction $i$ lies in the planes. Note that this effect
should also be present in layered superconductors with $s$ symmetry
within the layers. The magnitude of the $\int j_{\perp} j_{\parallel}$
coupling is proportional to
$\frac{\lambda_{\perp}^2 - \lambda_{\parallel}^2}{\lambda_{\perp}^2 
+ \lambda_{\parallel}^2}$.

\begin{figure}
\begin{center}
\mbox{\epsfxsize 8cm\epsfbox{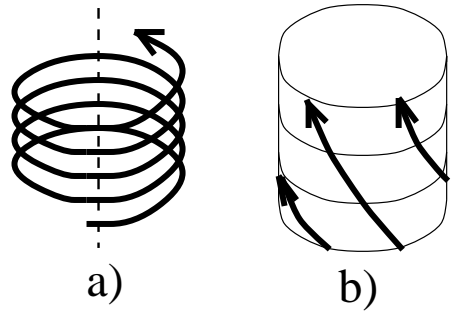}}
\end{center}
\caption{(a) Sketch of the currents around a vortex pinned
by a screw dislocation. Dashed line is the dislocation axis.
(b) Currents flowing along a twisted monocrystaline whisker.}
\label{dislocation}
\end{figure}

A particular defect which generates rotations in the 
out of the plane direction is a
screw dislocation,
which seems to be common in sputtered
films\cite{G91}. In cylindrical coordinates, we have that
only $\partial_{\theta} u_z = \frac{b_z}{2 \pi}$ is different 
from zero\cite{LL59}.  Hence, the free energy of the superconductor
has a term like $\int j_{\theta} j_z$. A current flowing around the
dislocation induces a current parallel to the axis of the dislocation.
This situation may arise if a vortex is pinned by a screw
dislocation. 
A similar effect can be
realized in a monocrystalline whisker subject to shear strains. 
Then, $\partial_z u_{\theta} \ne 0$, leading again to a 
$\int j_{\theta} j_z$ coupling. A current flowing
along the axis of the whisker induces a tangential component, giving rise
to a magnetic field in its interior.
A schematic picture of both situations is sketched in fig.(\ref{dislocation}).

\section{Conclusions.}
We have presented a formalism to study the influence of lattice rotations
on anisotropic superconductors. The analysis has been particularized 
to d-wave layered systems, but it is general enough to be
applicable to more complex lattices and order parameters.
We expect the scheme outlined here to be also useful for the
study of heavy fermions and other materials which show an
anisotropic order parameter.

For the systems considered here, we find  that

i) Lattice distortions couple directly to
an anisotropic order parameter, irrespective of other mechanisms,
like the dependence of T$_{\rm c}$ on pressure. 

ii) Higher
harmonics of the order parameter are more strongly frustrated
by local rotations\cite{KL65}. 

iii) Lattice defects, such as dislocations, reduce locally
T$_{\bf c}$. 
In the presence of a magnetic field, the vortex cores
will be attracted to these regions, giving rise to pinning.

iv) Transversely
polarized phonons also reduce T$_{\rm c}$. 
The value of the critical temperature
cannot be much larger than the typical frequency of these phonons.

v) Local rotations deflect supercurrents in superconductors
with inequivalent lattice axes. In particular, the current through
a grain boundary is deflected.

\section{Acknowledgements.}

This work was done with support from CICyT (Spain) through grant
PB96-0875. Valuable discussions with K. Maki, and hospitality at
the Istituto Eduardo Caianello (Vietri, Italy), are acknowledged.


\begin{references}
\bibitem{phononsw}
The present scheme cannot be used to analyze the pair breaking due
of short wavelength phonons. It will be studied elsewhere.
\bibitem{gradients}
These terms are discussed in detail in\cite{G89}.
\end{references}
\end{document}